\documentclass{openjournal}
\usepackage[utf8]{inputenc}
\usepackage[colorlinks,citecolor=blue,urlcolor=blue,bookmarks=false,hypertexnames=true]{hyperref} 

\begin{document}

\title{A SiPM photon-counting readout system for Ultra-Fast Astronomy}

\author{Albert Wai Kit Lau}
\affiliation{Department of Physics, The Hong Kong University of Science and Technology, Clear Water Bay, Kowloon, Hong Kong}

\author{Yan Yan Chan}
\affiliation{Department of Physics, The Hong Kong University of Science and Technology, Clear Water Bay, Kowloon, Hong Kong}

\author{Mehdi Shafiee}
\affiliation{Energetic Cosmos Laboratory, Nazarbayev University, Nur-sultan, Kazakhstan}
\affiliation{Department of Physics, Engineering Physics and Astronomy, Queen’s University, 64 Bader,Kingston, ON Canada}
\affiliation{Arthur B. McDonald Canadian Astroparticle Physics Research Institute, 64 Bader, Kingston, ON Canada}

\author{George F. Smoot,}
\affiliation{Energetic Cosmos Laboratory, Nazarbayev University, Nur-sultan, Kazakhstan}
\affiliation{Institute for Advanced Study Hong Kong University of Science and Technology, Clear Water Bay, Kowloon, Hong Kong} 
\affiliation{Université Sorbonne Paris Cité, Laboratoire APC-PCCP, Université Paris Diderot}
\affiliation{Department of Physics, University of California, California, USA }
\affiliation{Universit{\'e}  de  Paris, Laboratoire Astroparticule  et  Cosmologie,  F-75013
Paris, France} 
\affiliation{Donostia International Physics Center,  University of the Basque  Country  UPV/EHU,  E-48080  San  Sebastian,  Spain}

\begin{abstract}
Very little work has been done searching for astrophysical transient optical emission in the millisecond to nanosecond regime with significant sensitivity. 
We call this regime “Ultra-Fast Astronomy”, or UFA. To investigate transients on as short time scales as possible, we developed our own customized readout system for a silicon photomultiplier (SiPM)-based UFA camera, intended for use on conventional astronomical telescopes. SiPMs, available in array packages for imaging a field, are capable of time-tagged single-photon detection in the visible wavelength range. Our readout system consists of 16 channels of 14-bit data logging. Each channel includes a 50-dB gain pre-amplifier, signal shaping circuits, an analogue front end, an analogue to digital converter, and a Xilinx UltraScale+ Field Programable Gate Array Multipurpose System on Chip (FPGA-MPSoC) board for data-logging. We show that our system successfully read out the data from SiPM at 16 ns intervals with a maximum power consumption of 300 mW per channel and capability to perform concurrent 16 channels readout.
\end{abstract}

\author{Bruce Grossan}
\affiliation{Energetic Cosmos Laboratory, Nazarbayev University, Nur-sultan, Kazakhstan}
\affiliation{UC Berkeley Space Sciences Laboratory, California, USA}


\date{November 2021}

\maketitle

\section{Introduction}
\label{sec:intro}
    Silicon Photomultipliers (SiPMs) are a type of solid-state photon-counting Geiger-mode avalanche diode offered commercially. 
    These devices are gaining popularity in scientific research due to their robustness, immunity to magnetic fields, lack of high-voltage, and excellent photon resolving ability compared to photomultiplier tubes (PMTs). The output from a SiPM is charge pulses of 100 ns time scale, and the total charge within the pulse is proportional to the detected photon counts arriving within this time. In our Ultra-Fast Astronomy (UFA) Program, we wish to develop a SiPM-based camera to explore astronomy at time scales shorter than milliseconds which is not easily accessible via traditional CCD or CMOS sensors \citep{li2019program, denissenya2021ultra}. 
    
    For most SiPM readout systems, application-specific integrated circuits (ASICs) are used to provide timing down to picosecond (ps) accuracy. However, such ASICs usually adopt a peak-locking architecture for reading both timing and charge information \citep{auger2016multi}. Peak-locking architecture freezes the readout system when a photon triggers the circuit, introducing a restoring dead time of typically microseconds after each trigger. For UFA, we would like to retrieve information from every detected photon, so the dead time should be eliminated \citep{lau2020sky, Shafiee_2021}. Also, the system should be low-cost to allow scaling up to large arrays in the future. Here we summarise the main requirements of the system:
    \begin{itemize}
        \item 
            16-channel SiPM readout on a single readout board
        \item Low development and production cost of $<40$ USD per channel
        
        
        \item Working temperature of $-40^\circ C$ to $85^\circ C$ for harsh mountain observatory conditions
        \item $< 20ns$ sampling time interval for photon arrival time recording
        \item Discrimination of SiPM signal from 0 to 
            10+ SiPM photoelectrons
        \item No readout-induced dead time
        \item On-board data processing system 
    \end{itemize}
    
    
    From the above requirements, we designed a SiPM readout system consisting of a front-end board made from commercially available integrated circuit (IC) components, and a board to format and further process digitized pulses, an entry-level Field Programable Gate Array Multipurpose System on Chip (FPGA-MPSoC). We describe these two boards in more detail below.

\section{Front-end board design}
\subsection{Overview and supporting circuits}
    Our front-end board incorporates two stages of pre-amplifiers (using an INA-03184 and a PGA5807a), a customized signal shaper 
    based on an OPA4820, and ADC readout based on an AFE5818. We choose the pre-amplifiers to provide minimal noise with reasonable circuit complexity and power requirements. The OPA4820 used in the signal shaper provides a high gain-bandwidth product to ensure linearity at high frequency, and the AFE5818 ADC provides maximum flexibility via a high-speed ADC readout with a programmable gain.
    
\begin{figure}[h!]
        \centering
        \includegraphics[width = \textwidth]{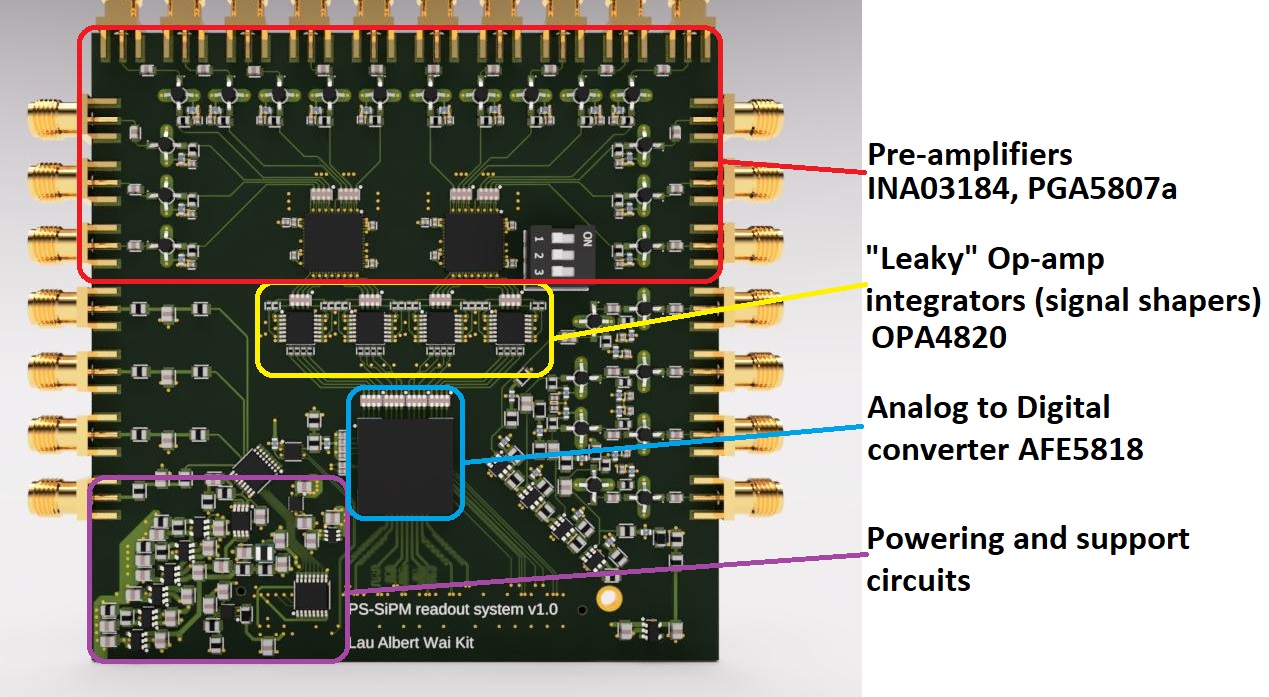}
        \caption{Rendering of the front-end board PCB design. The major circuit components and functional component groups are indicated.}
        \label{fig:pcb_render}
    \end{figure}

    
    This front-end board is designed on a single four-layer PCB with impedance control to ensure low-cost PCB fabrication.  A rendering of the PCB design is shown in Fig.\ref{fig:pcb_render}
    , with components grouped by function. Ungrouped components are currently unused. 

    The system obtains +12V and +3.3V supplies from FPGA Mezzanine Card Low Pin Count (FMC-LPC) interface, which are supplied to the different ICs through the conversion paths given in Table \ref{tab:power_grouping}.
    
    \begin{table}[h!]
        \centering
        \caption{Power supplies conversion path of different ICs}
        \begin{tabular}{||c||c|c|c||}
        \hline
        $V_{IN}$ & Function grouping & Conversion path & $V_{OUT}$ \\
        \hline
        12V & Pre-amplifier INA-03184 & TPS82130 $\rightarrow$ TPS7A20 & 5V\\
        \hline
        3.3V & Pre-amplifier PGA5807A & TLV758P & 3.2V\\
        \hline
        12V & Signal Shaper & TPS82130 $\rightarrow$ TPS7A20 & 5V\\
        \hline 
        12V & & TPS82130 $\rightarrow$ TPS7A20 & 5V\\
        3.3V & AFE5818 analog power & TLV758P & 3.2V\\
        1.8V & & TPS73601 & 1.75V\\
        \hline
        1.8V & AFE5818 digital power& LP5912 & 1.2V\\
        1.8V & & TPS22919 & 1.8V\\
        \hline
        \end{tabular}
        \label{tab:power_grouping}
    \end{table}

\subsection{Pre-amplifiers}
    The charge pulses from SiPM are first converted to voltage pulses via a $50\Omega$ resistor and 
    are transmitted through $50\Omega$ coupling SMA cables to the readout board. The signal is then amplified by two-stages of pre-amplifiers, a low-noise amplifier INA-03184 followed by an PGA5807A amplifier, on the readout board. The INA-03184, from Hewlett Packard, is a silicon bipolar monolithic microwave integrated circuit amplifier with a low noise figure (2.5dB),  a low power consumption (50mW per channel) and a simple circuit. It provides a flat 26dB gain up to 2GHz \citep{hpamp_03184}.
    
    The second stage pre-amplifier is a Texas Instruments PGA5807A, with eight channels tunable from 12dB to 30dB gain up to 75MHz. This amplifier has a low pass filter active above 75MHz, eliminating thermal noise at higher frequencies. The power consumption of this amplifier is 60mW per channel. Since the PGA5807A is designed to work on differential signals, while we have only a single signal channel, half of the signal amplitude (6dB) is lost \citep{PGA5807a}. From this amplifier chain, the SiPM signal can be amplified by at most 50dB. The signal will then be fed into the signal shaper for further processing.

\subsection{The signal shaper}
    A signal shaper performs analogue integration of the pulse (and the result is proportional to 
    the charge in 
    a SiPM pulse). This component replaces a high-speed digital integration algorithm in the FPGA downstream.
    
\begin{figure}[h!]
        \centering
        \includegraphics[width = \textwidth]{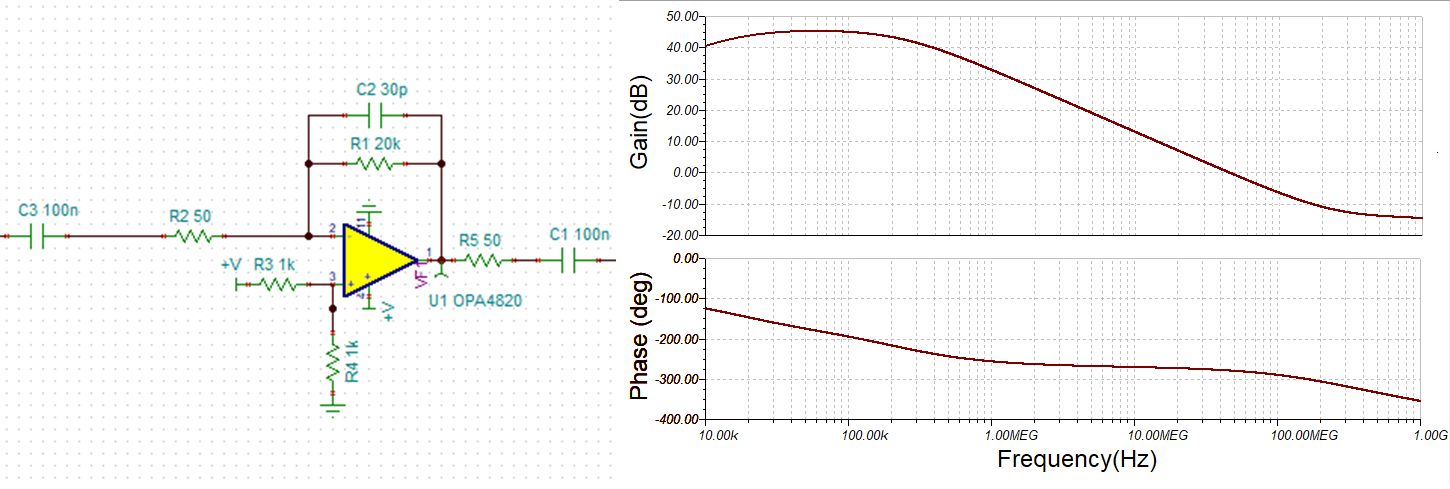}
        \caption{Circuit schematic and simulated AC transfer characteristic of signal shaper}
        \label{fig:integrator}
    \end{figure}

    We use a "leaky" integrator design for our shaper, 
    based on an OPA4820 op-amp IC. the OPA4820 provides four channels of high speed (650MHz gain-bandwidth product) amplifiers on a single rail 5V power supply \citep{OPA4820}. The shaper consists of an op-amp integrator, with negative feedback from a $30pF$ capacitor and a $20k\Omega$ resistor on  as shown in Fig. \ref{fig:integrator}. The resistor in the feedback line causes the signal to gradually decay with a time constant of $\sim 600ns$. Such a "leaky" characteristic ensures 
    that the shaper will not encounter saturation. Compared to a traditional op-amp integrator with a reset switch, the "leaky" design ensures no dead-time signal shaping.
   
    The circuit is simulated using the Simulation Program with Integrated Circuit Emphasis (SPICE) program TINA-TI, as shown in Fig. \ref{fig:integrator}. The simulation result (AC Bode plot) shows that the shaper has an integration band from 1MHz to 100MHz (gain plot slope -20dB per decade) without ring features. 
    The plot also shows that the shaper provides an additional gain of $\sim 10dB$ on a 10MHz signal, which is the typical frequency of most SiPM pulses.
    
    The OPA4820 consumes at most 6mA quiescent current per channel with a 5V supply, with maximum power consumption of 50mW under a large signal input. 
    

\subsection{Analog to Digital conversion}
    After shaping, the analogue signal is ready for digital conversion, with a signal on the order of $mV$.  
    This is achieved using a medical analogue front end IC AFE5818 from Texas Instruments. The AFE5818 has 16 channels of 14bit analogue to digital conversion (ADC), at 65 mega samples per second (Msps) \citep{AFE5818}. However, due to the limitation of the FPGA Low Voltage Differential Signal (LVDS) interface limitation (16bit serialization, 1Gbps maximum), we set the ADC speed to 62.5Msps. 

    The AFE5818 also performs internal signal amplification and shaping, 
    providing an internal gain tunable from -4 to 54dB. In addition, the
    AFE5818 has a tunable 3rd-order, linear-phase low pass filter from 10MHz to 50MHz and a high pass filter tunable from 15kHz to 200kHz, allowing the removal of unwanted high-frequency noise and baseline fluctuation. The power consumption of the AFE5818 is 140mW per channel maximum.

\section{FPGA+ARM MPSoC-based digital processor system design}

\begin{figure}[h!]
        \centering
        \includegraphics[width = 0.8\textwidth]{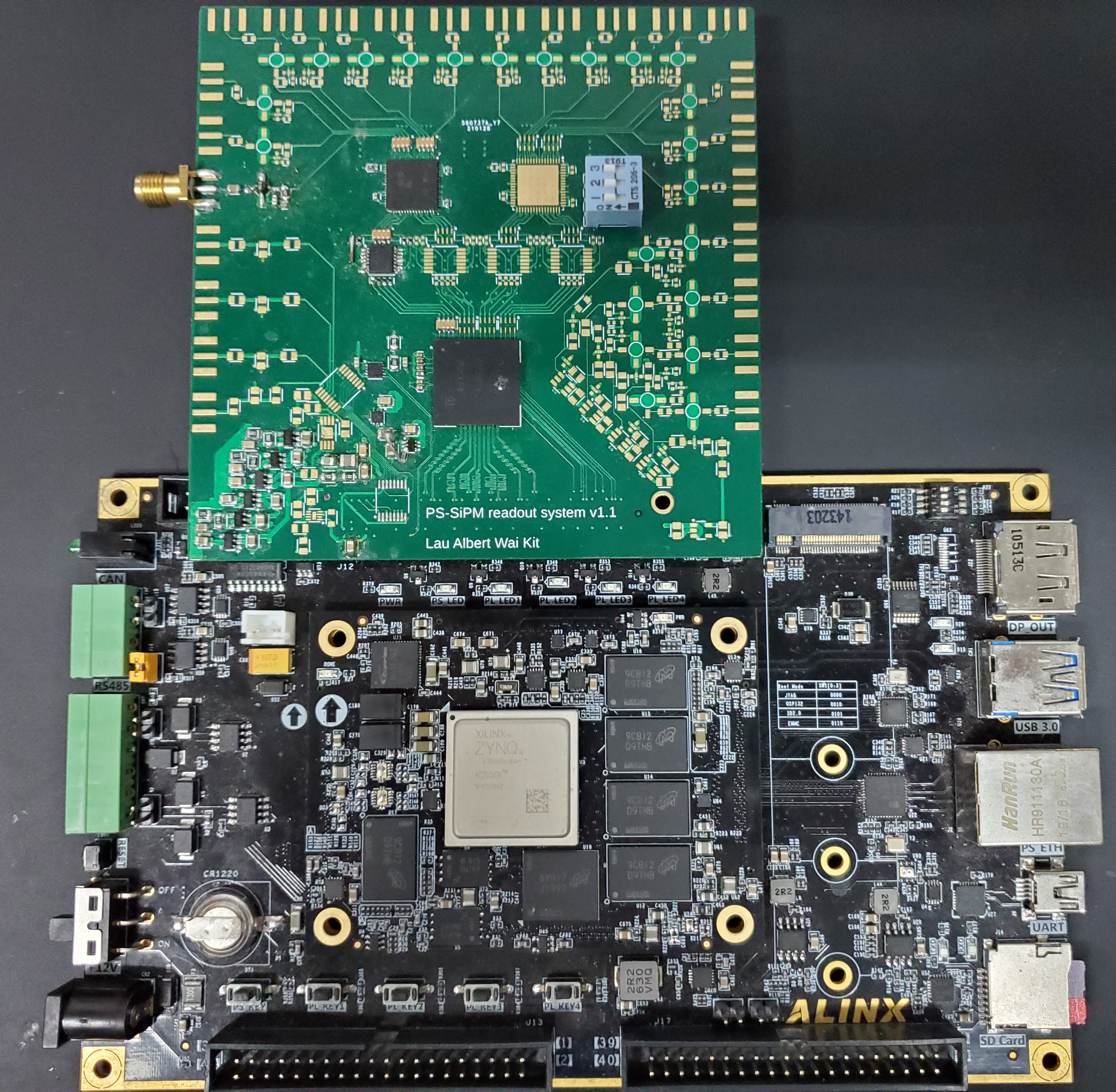}
        \caption{Image of the FPGA development board with readout system (partly populated)}
        \label{fig:FPGA}
    \end{figure}

    We selected a Xilinx UltraScale+ MPSoC (model xczu3cg-1sfvc784e) development board from Alinx Inc. for processing the digital data coming from the ADC. \citep{boppana2015ultrascale+}. First, the LVDS output from the AFE5818 ADC passes through FPGA Mezzanine Card Low Pin Count (FMC-LPC) interface to the high-performance I/O pins of the FPGA. The FPGA then 
    performs deserialization and 16-bit word alignment to obtain ADC digitalized data. Finally, the  ADC digitalized data are transmitted to the ARM core running a petalinux system through an Advanced eXtensible Interface 4 Direct Memory Access (AXI4-DMA) with a first-in first-out (FIFO) buffer. The petalinux system running on ARM core format the data and prepare them for post-processing. An image of the system (without heat-sinks, single channel equiped) is shown in Fig. \ref{fig:FPGA}. 
    
\begin{figure}[h!]
        \centering
        \includegraphics[width = \textwidth]{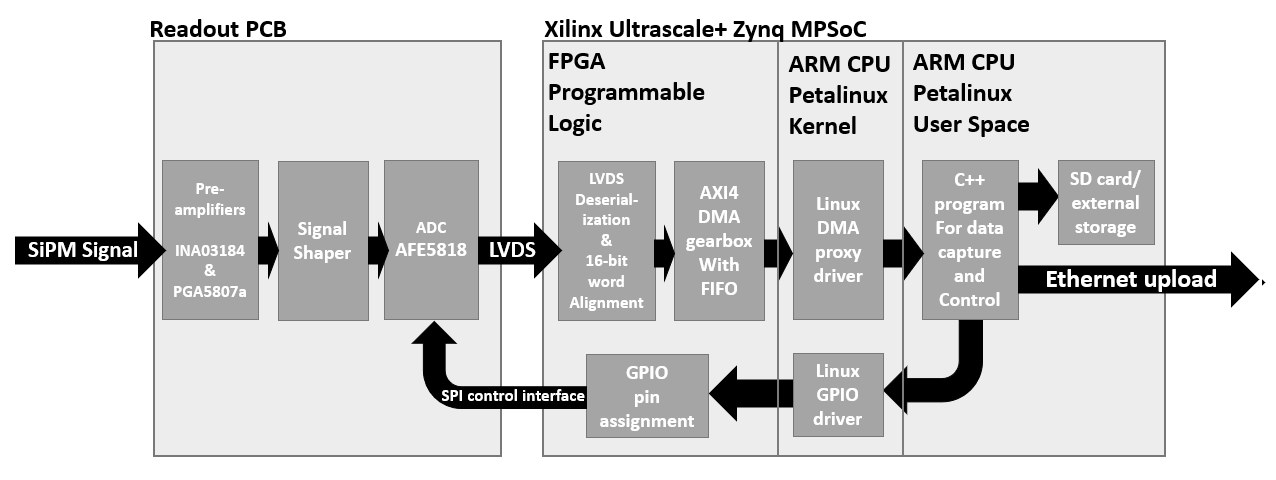}
        \caption{Data-flow and control diagram of the readout system}
        \label{fig:FPGA_block_diagram}
    \end{figure}

   A diagram of the data flow and control is shown in Fig. \ref{fig:FPGA_block_diagram}. Collected data 
    are sent to a host computer via a Gigabit ethernet connection.
    
    \section{SiPM signal experimental results}
    We assembled the readout system described above, and connected it to a Hamamatsu S14520-3050VS SiPM for testing. The detector was kept at room temperature, $25 \pm 0.5^\circ C$ on a metal block as heat reservoir and monitored by a NTC-10k thermistor. The breakdown voltage of
    the SiPM was measured to be $39.47V$ with 
    the overvoltage kept at $3V$, i.e. the power supply voltage $=39.47+3 = 42.47V$. The manufacturer specifications for these conditions give a SiPM
photon gain of $2.8\times10^6$ and a dark count of $600kcps$ (kilo-counts per second) \citep{yamamoto2019recent}. The temperature coefficient of breakdown voltage is $34mV/^\circ C$. We can safely assume the thermal fluctuation of SiPM gain is $\pm0.5\times0.034/3 \sim \pm 0.5\%$ with gain linear proportional to overvoltage.

    The internal gain of the AFE5818 ADC was set to 0dB, the high pass filter frequency to 150kHz, and low pass filter frequency to 50MHz. 

\begin{figure}[h!]
        \centering
        \includegraphics[width=0.9\textwidth]{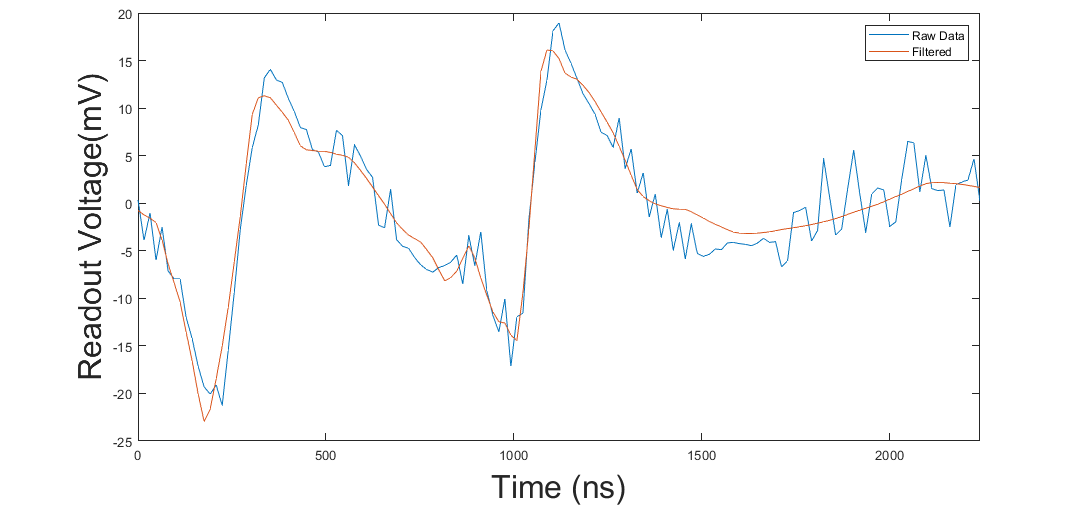}
        \caption{Dark count signal captured by SiPM readout system}
        \label{fig:ADC_1pe}
    \end{figure}

    The system was placed in a dark grounded Faraday cage made with aluminum board, and dark counts of the SiPM 
   were measured through the readout system, as shown in Fig.\ref{fig:ADC_1pe}. 
    The rise time of each count is about $\sim 50ns$. Taking trigger on $\frac{1}{3}$ p.e. level (10mV) and assuming a linear rising, the time uncertainty from signal arrival to trigger will be less than 20ns. With a 16ns sampling interval, the final photon arrival time logging accuracy of $\pm$16ns can be achieved. 
    
    
    The decay tail of photon pulses are $\sim500ns$ long, corresponding to the time constant of the signal shaper. The 
    height of the recorded readout pulses 
 should be linear with the charge pulse within the SiPM
    (yielding about $30 mV$ / photoelectron),
    because of the properties of the integrator. 
  The number of concurrent incoming photons can therefore be 
  determined by the readout pulse height. Since the ADC 
    has a full range of $\pm 1V$, it is possible to detect more than 30 concurrent photoelectrons without saturation. 
    
    
\begin{figure}[h!]
        \centering
        \includegraphics[width=\textwidth]{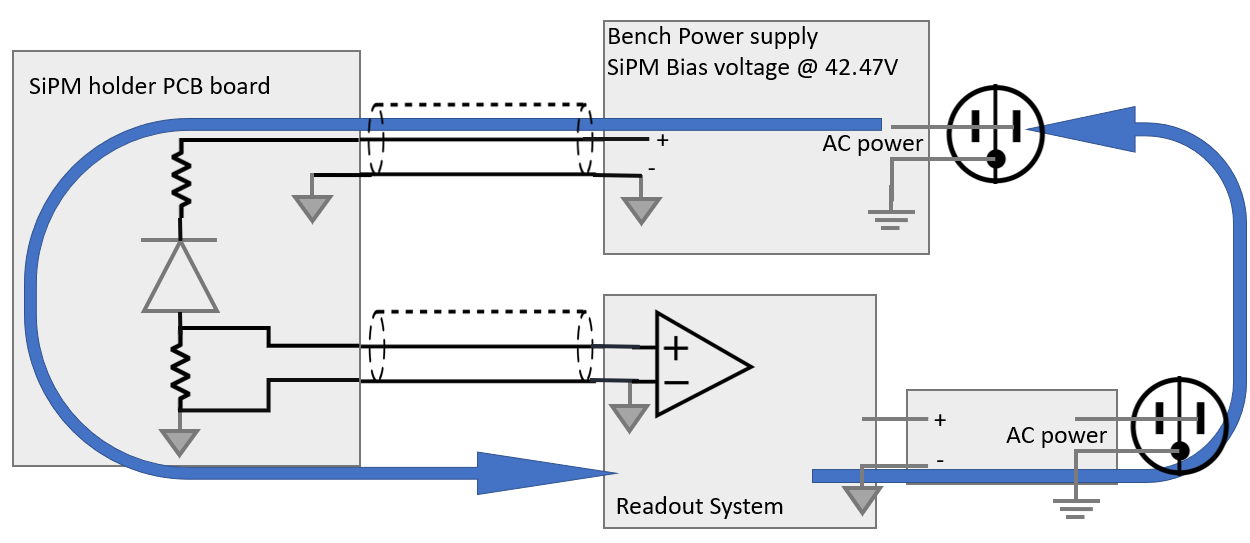}
        \caption{Schematic of testing setup. The proposed ground loop path is indicated by the blue arrow.}
        \label{fig:ground_loop}
    \end{figure}

    From the figure, some noise can be seen on the SiPM pulse. This noise 
    could possibly come from a ground loop from an external power supply, through the SiPM board to the readout system, as indicated in Fig. \ref{fig:ground_loop}.  To remove this noise, we implemented a Symlets 4 wavelet filter in Matlab \citep{chavan2011implementation, Shafiee_2016}
    as shown by the reddish line in Fig. \ref{fig:ADC_1pe}.

\begin{figure}[h!]
        \centering
        \includegraphics[width=\textwidth]{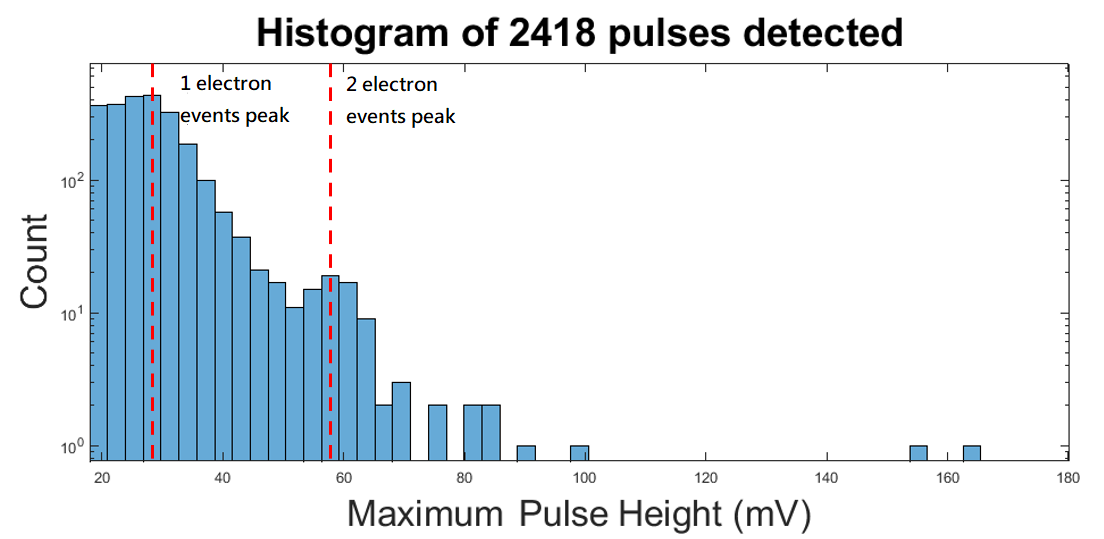}
        \caption{Pulse height distribution plot of dark counts.}
        \label{fig:hist_log}
    \end{figure}

    Limited by the single-pass AXI4-DMA transfer size of $512kB$,
    we run the readout system for $16ns*512*1024/2 = 4.194ms$ (each sample takes 2 bytes, 16ns sampling time).  
    This limit will be eliminated in the future by adding a FIFO buffer based on DDR4 memory. In total, 2184 dark counts were detected with a detection threshold of $15mV$, indicating a darkcount rate of $~520kcps$, similar to the manufacturer specification of $600kcps$. 
    The pulse height distribution is plotted as a histogram in Fig. \ref{fig:hist_log}
    .The plot shows two noticeable peaks: the first one (corresponding to a single electron) at $\sim$30mV and the second one (two electrons) at $\sim$60mV. The second peak has $\sim 5\%$ count of the first one, matching the $5\%$ crosstalk specification of S14520-3050VS SiPM \citep{yamamoto2019recent}. This demonstrates the system's ability to distinguish photon counts from the output signal from the SiPM.

\section{Summary and Future Work}

    A novel, low-cost SiPM readout system has been developed for the 
    UFA project. The readout system incorporates analogue processing and digitization ICs, data collection through FPGA and communication to host computer through Ethernet linkage. The readout system provides SiPM signal logging with $\pm$16ns photon arrival time logging accuracy, and photon counts are recovered in post-processing algorithm. 
    
    The readout board (without FPGA) takes at most $50+60+50+140 = 300mW$ per channel with a per-channel cost of less than 20USD. The readout board is built with industrial standards (working temperature from $-40^\circ C$ to $85^\circ C$) for harsh weather at the observatory. 
    
    
    
    
    We plan to build on our success in making a fast, low-cost per channel system by decreasing noise, increasing readout capabilities, and improving imaging.  In order to accomplish this, we plan to begin work immediately on the following tasks: 
    
    \begin{itemize}
        \item Move the SiPM powering circuit onto the front-end PCB to reduce ground loop noise
        \item Simplify the analog amplifier chain by utilizing the internal amplifier in the AFE5818
        \item Perform low-temperature testing in the laboratory and replace failed board and components for upcoming field tests
        \item Add a DDR4 memory-based FIFO buffer for 
        continuous data collection
        \item Integrate photon count recovery post-processing algorithm into FPGA logic
        \item Implement imaging by using a position-sensitive SiPM and a  position-decoding algorithm within the FPGA.
    \end{itemize}
    
    
    The goal of  this work is to produce an astronomical camera that can be used for transient searches on all $\sim$ sub-second time scales down to our limiting time scale, and for searches for high frequency non-random signal patterns, from ultra-short periodicity to SETI signals. With the additional work listed, the readout system should be fully capable of supporting this goal.

\acknowledgments
We acknowledge 
the support from the HKUST Jockey Club Institute for Advanced Study (IAS).

\bibliography{reference}

\end{document}